\documentclass[twocolumn,showpacs,amsmath,amssymb]{revtex4}
\usepackage{graphicx}
\usepackage{dcolumn}
\usepackage{bm}

\newcommand{\ba}{\begin{eqnarray}}
\newcommand{\ea}{\end{eqnarray}}
\newcommand{\be}{\begin{equation}}
\newcommand{\ee}{\end{equation}}
\newcommand{\bd}{\begin{displaymath}}
\newcommand{\ed}{\end{displaymath}}

\begin{document}

\draft


\title{Entanglement Entropy in Critical Phenomena and
Analogue Models of Quantum Gravity}
\author{Dmitri V. Fursaev}
\address{
  \medskip
  Dubna International University and\\
  University Centre\\
  Joint Institute for Nuclear Research\\
  141 980, Dubna, Moscow Region, Russia\\
  {\rm E-mail: \texttt{fursaev@thsun1.jinr.ru}}\\
  \medskip
}

\begin{abstract}
A general geometrical structure of the entanglement entropy for spatial
partition of a relativistic QFT system is established by
using methods  of the effective
gravity action and the spectral geometry. A special attention is payed to
the subleading terms in the entropy in different dimensions and to behaviour
in different states. It is conjectured,
on the base of relation between
the entropy and the action, that in a fundamental theory the
ground state entanglement entropy  per unit area equals
$1/(4G_N)$, where $G_N$ is the Newton constant in the low-energy
gravity sector of the theory. The conjecture opens a new avenue
in analogue gravity models. For instance, in higher dimensional condensed
matter systems, which near a critical point are described by
relativistic QFT's, the entanglement entropy
density defines an effective gravitational coupling. By studying
the properties of this constant one can get new insights in
quantum gravity phenomena, such as  the universality of the
low-energy physics, the renormalization group behavior of $G_N$,
the statistical meaning of the Bekenstein-Hawking entropy.
\end{abstract}


\maketitle

\newpage

\section{Fundamental entanglement}
\subsection{General properties}

Quantum entanglement is an important physical phenomenon
in which the quantum states of several objects cannot be described
independently, even if the objects are spatially separated.
Quantum entanglement is used in different
research areas \cite{Pres}.
In quantum information theory entangled states are a valuable
source of processing  the information. Quantum entanglement also plays an important
role in properties of strongly correlated many-body systems and in
collective phenomena  such as quantum phase transitions. In quantum gravity
the entanglement may be a key for understanding the mystery of the
black hole entropy.

The entanglement can be quantified by an entropy. One can
define it as the measure of the information
about quantum states which is lost when these states cannot be observed.
In many-body systems, which are the subject of the
present work,  "observable" and "unobservable" states can be located in
different regions.
Consider, for instance,  a
lattice of spins being in a quantum state characterized by a density matrix
$\hat{\rho}$. Suppose that the lattice is divided into regions $A$
and $B$ with a common boundary $\cal B$. The entanglement between
the two regions can be described by the reduced density matrix
$$
\hat{{\rho}}_B=\mbox{Tr}_A \hat{\rho}~~,
$$
where the trace is taken over
the states of spin operators at the lattice sites in the
 region $A$. The entanglement entropy in the region $B$
is defined as the von Neumann entropy
\begin{equation}\label{ent}
S_B=-\mbox{Tr}_B \hat{{\rho}}_B\ln \hat{{\rho}}_B~~.
\end{equation}
Analogously one can define the entanglement entropy $S_A$ in the region $A$ by tracing
the density matrix over the states located in the region $B$.
If the system is in a pure state $|\psi\rangle$, i.e. $\hat{{\rho}}=|\psi\rangle\langle\psi |$,
it is not difficult to show that $S_A=S_B$, see \cite{Sr:93}.

It has been observed long ago \cite{Sr:93},\cite{BKLS}
(see also recent works \cite{Plenio1},
\cite{Plenio2}), that in lattice models representing a discrete
version of a relativistic quantum field theory (QFT) the ground state
entanglement entropy in the leading order is proportional to the
area
$\cal A$ of $\cal B$,
\begin{equation}\label{i1}
S=S_B \sim {{\cal A} \over \varrho^{D-2}}~~.
\end{equation}
Here $\varrho$
is the lattice spacing and
$D$ is the number of spacetimes dimensions (it is assumed that $D>2$).
This geometrical property of  $S$
follows from the fact that field excitations in different regions are
correlated across the boundary.

\subsection{Black hole entropy}

If $\varrho$ is identified with the Planck length $l_{Pl}$,
(\ref{i1}) looks very similar to the Bekenstein-Hawking entropy of a black hole
\begin{equation}\label{i2}
S^{BH} = {{\cal A}_H \over 4G_N}~~.
\end{equation}
Here ${\cal A}_H$ is the area surface of the black hole horizon and $G_N$ is the Newton
constant, $G_N=l_{Pl}^{D-2}$. The similarity of the two entropies suggests that $S^{BH}$
may be an entanglement entropy, as was first pointed out in \cite{Sr:93},\cite{BKLS}. An external
observer, who is at rest with
respect to a black hole, never
sees what happens inside the horizon. Because the observer perceives the vacuum excitations
in a mixed state it is natural to assume that $S^{BH}$ measures
the loss of the information inside the horizon.

The reduced density matrix for a black hole is thermal \cite{Israel:76}. Thus, the entanglement entropy
coincides with the entropy of a thermal atmosphere around the black hole horizon
(see \cite{FrFu:98} for a review of different interpretations of $S^{BH}$ in terms
of the entropy of quantum excitations).
However, relation between $S^{BH}$ and the entropy of entanglement (or thermal
entropy) is non-trivial because $S$ is divergent. In the lattice
regularization (\ref{i1}) diverges in the continuum limit when $\varrho \rightarrow 0$.
It was pointed out in \cite{SuUg:94},
\cite{CaWi:94} that the divergences in $S$ are the standard ultraviolet
divergences. If the entropy of thermal atmosphere
is considered as a quantum correction to $S^{BH}$
the divergences  can be eliminated in the course
of renormalization of the gravitational
coupling.

A deep connection
between the divergences in $S$ and the divergences in the quantum effective action
has been established latter for different field species at the one-loop level \cite{FrFu:98}.
This connection is very important. Suppose, following the
original idea by Sakharov  \cite{Sakh}, that the Einstein action is entirely
induced by quantum effects
of some underlying fundamental theory. Because the Newton constant in this case has
a pure quantum origin, so does the entropy $S^{BH}$.
The hypothesis is that the entropy of a black hole in such a theory is
an entanglement entropy of fundamental quantum excitations which induce
low-energy gravity dynamics \cite{Jaco:94}--\cite{FF:97}.

The mechanism of generation of the black hole entropy in induced gravity may be quite
general and, in particular, it can be realized in string theory \cite{MHS}.
In string theory the low-energy effective gravity action appears from the tree-level
diagrams of closed strings. From the point of view of open strings
these are one-loop diagrams. Thus, the gravity action as well as the Bekenstein-Hawking
entropy can be interpreted as a pure loop effect. Another argument in favour of
entanglement interpretation of black hole entropy comes from analysis
of string theory on asymptotically anti-de Sitter backgrounds and its
duality to a conformal field theory (CFT) \cite{Mald},\cite{BEY}.

\subsection{Entanglement entropy in fundamental theories}

The purpose of the present work is to look at these results from a different
perspective. Suppose the underlying fundamental theory of quantum gravity  is known.
This can be a string--$D$-brane theory or some other theory which correctly describes
the observable particle physics and gravitational effects.
Consider a low-energy
limit of this theory and the entanglement entropy $S$ in the ground state
for partition of the system in Minkowsky spacetime by a plane.
It is assumed that $S$ measures entanglement
of all genuine microscopical degrees of freedom of the fundamental theory
(which may not coincide with the low-energy fields).
We call such quantity the {\it fundamental entanglement entropy}.

Because the space is infinite it is convenient
to discuss the density of $S$ per unit area. Let us denote this quantity by ${\cal S}$.
We make a conjecture that the density of the fundamental entanglement entropy
 is finite and is given precisely by the relation
\begin{equation}\label{i3}
{\cal S} = {1 \over 4G_N}~~,
\end{equation}
where $G_N$ is the Newton constant in the low-energy gravity sector of the theory.
The conjecture is based on the above mentioned arguments:

i) in a fundamental theory the cutoff parameter $\varrho$ in (\ref{i1}) should be of the order
of the Planck length;

ii) there are evidences that the Bekenstein-Hawking entropy  (\ref{i2})
is the entanglement entropy for partition of the system by the horizon.

In what follows we discuss a more formal argument  in favour of (\ref{i3}).
It is based on the observation that the black hole entropy and
the entropy of entanglement are derived in a unique way from the
effective gravity action.

\bigskip

Relation (\ref{i3}) has a number of interesting consequences. One of them is that
the definition  of $G_N$ becomes not only the subject
of gravitational physics. The flat space entanglement is an alternative source
of information about the Newton constant. (In a certain sense the situation
with different definitions of $G_N$  can be compared with "gravitational" and the
"inertional" notions of the mass.)

In this paper we pay attention to the fact that
relation (\ref{i3}) can be used to study the properties
of the gravitational coupling by carrying the computations of the
entanglement entropy in condensed matter systems.
There are higher-dimensional condensed matter systems which near a critical point corresponding to a
second order phase transition are described by relativistic QFT's which contain
massive fields. The class of Ising models is one of such examples. The
QFT's in these systems are analogous to low-energy limit of the fundamental theory.
The relativistic covariance guarantees that the entanglement entropy density
in the near-critical regime determines an effective gravitational coupling by
the same equation as Eq. (\ref{i3}).
The lattice spacing in the underlying theory serves as a cutoff. Hence
the entanglement entropy is finite and can be calculated by using analytical and numerical
methods. This enables one to pose different questions about properties of $G_N$ and try to find
the answers in gravity analogs.

\subsection{Outline of the paper}

The rest part of the paper is organized as follows.
Section II is devoted to the entanglement entropy in relativistic QFT's.
The Section starts with discussion of one-dimensional spin models which serve as an
illustration.
We then  establish connection
between the entanglement entropy $S$ in relativistic
QFT's and the effective gravity action, and use methods
of the spectral geometry for derivation of $S$.
By studying the theories in a cubic
region we find a geometric form of $S$ in different dimensions, at zero and high
temperatures. We pay a special attention at boundary effects. Our result for
the ground state entanglement
in four-dimensional space-times looks as follows:
\begin{equation}\label{main}
S\sim {{\cal A} \over \varrho^2} +{{\cal L} \over \varrho}+ {\cal C}\ln \varrho ~~~,
\end{equation}
where $\cal A$ is the area of the separating surface $\cal B$ and $\cal L$ is the
length of the boundary of $\cal B$. The last term in $S$ has a topological form. It depends on
the number $\cal C$ of sharp corners (vertexes) of the boundary of $\cal B$ and does not
change under smooth deformations of $\cal B$. Calculation of the subleading terms in
(\ref{main}) is the new result.

Relation of $S$ to  the Bekenstein-Hawking entropy in induced gravity models in four
dimensions is discussed in Section III. We also discuss the concept
of gravity analogs and possible
models which can be used to study properties of the gravitational coupling on the base
of relation (\ref{i3}). There are several topics which can be pursued by using
simulations in the
new class of gravity analogs. Some of them such as the renormalization group behavior of $G_N$,
the universality of the low-energy physics and the statistical meaning
of the Bekenstein-Hawking entropy are listed in Section IV.

\section{Entanglement in relativistic field theories}
\subsection{One-dimensional spin chains}

We begin with discussion of the properties of the entanglement entropy in
one-dimensional spin chains. Consider as a simplest example the entanglement entropy
for a block of
spins in the Ising model. The Hamiltonian of the model is
\begin{equation}\label{6}
H=\sum_{i=1}^N\left(\sigma_i^x\sigma_{i+1}^x+\lambda
\sigma_i^z\right)~~~,
\end{equation}
where $N$ is the number of spins and  $\sigma_i^x, \sigma_i^z$ are
the Pauli matrices. Parameter $\lambda$ is the strength of
external magnetic field. At zero temperature the Ising chain has a
second order phase transition at the critical value $\lambda=1$.

As is known, there is a unitary transformation (the Jordan-Wigner transformation) which
maps the spin models like (\ref{6}) to one-dimensional models with two spinless
fermions. This transformation can be used to diagonalize the Hamiltonian.

The entanglement entropy can be investigated for the ground state when the chain is separated
into two blocks of contiguous spins of equal sizes $N/2$. Suppose
that $\lambda$  is fixed and $N$ varies ($N$ is much larger than the correlation length).
The behaviour of $S$ has been studied in two regimes \cite{crit1}.
In the  off-critical regime, $\lambda\neq 1$ and
$|\lambda-1|\ll 1$, the entropy at large $N$ reaches the
saturation value
\begin{equation}\label{r1}
S(N,\lambda)= -\frac 16 \log_2 |\lambda-1|~~.
\end{equation}
In the critical regime, $\lambda=1$, the entropy does not reach
the saturation and behaves at large $N$ as
\begin{equation}\label{r2}
S(N,\lambda)\simeq \frac 16 \log_2 N/2~~~.
\end{equation}
The explanation of these results  is based on the fact
that in the
continuous limit, $N\rightarrow \infty$,
the Ising model near the critical point
corresponds to a quantum field theory with two fermion fields whose
mass $m$ (or the inverse correlation length $\xi^{-1}$) is monotonically related to
$|\lambda-1|$. In
the critical regime $m$ vanishes and one has a
conformal field theory  with two massless fermions each having
the central charge $1/2$.

Formulae (\ref{r1}), (\ref{r2}) are quite general. They have been verified in other
exactly solvable one-dimensional statistical models (XXZ, XY and other models)
which generalize the Izing chain (\ref{6}), see \cite{crit1}--\cite{crit4} and a review
of results in \cite{Rico}.
To discuss the continuum limit in lattice models
it is convenient to introduce the lattice spacing $\varrho$. Then the number of spins $N$ defines
the size of the system $L=N\varrho$. In the systems separated into two equal
parts  the entropy at the critical points behaves
as
\begin{equation}\label{r3}
S\simeq \frac c6 \ln {L \over 2\varrho}~~,
\end{equation}
where $c$ is a total central charge of the corresponding effective
two-dimensional conformal theory.
Near the critical point the entropy behaves
as
\begin{equation}\label{r4}
S\simeq \frac c6  \ln {1 \over \varrho m}~~,
\end{equation}
where $m$ is the corresponding inverse correlation length.
From the point of view of QFT $\varrho$ is the ultraviolet cutoff parameter.

\subsection{Entanglement entropy and effective gravity action}

There are different ways to derive (\ref{r3}), (\ref{r4}) by QFT methods.
Our purpose is to show how it can be done by using
the effective action approach. We first start with a theory in a flat spacetime with
arbitrary number of dimensions $D$ and then focus on theories in $D=2,3$, and 4.

The entanglement entropy (\ref{ent}) for a lattice system discussed in Section I.A can
be rewritten
as
\begin{equation}\label{11}
S_B=-\lim_{n\rightarrow 1}~{\partial \over \partial n}\mbox{Tr}_B
~\hat{{\rho}}_B^{~n} ~~~.
\end{equation}
Suppose that near a critical point the system is equivalent to a QFT
with field variables $\phi$ whose dynamics is determined by the action $I_E[\phi]$.
The density matrix $\hat{{\rho}}_B$
in the configuration representation depends on  variables $\phi_B$
in the region $B$.
We consider the system at a finite temperature $T$.
The result for the ground state entanglement
can be obtained in the limit $T\rightarrow 0$.
The matrix elements of $\hat{{\rho}}_B$ can
be described in terms of the Euclidean path integral
\begin{equation}\label{12}
\langle \phi'_B|\hat{{\rho}}_B|\phi_B\rangle = {\cal N}^{-1}\int^{\phi=\phi'_B}_{\phi=\phi_B}
[D\tilde{\phi}]~e^{-I_E[\phi]}~~~,
\end{equation}
where $\cal N$ is a normalization coefficient introduced to satisfy the condition $\mbox{Tr}_B
~\hat{{\rho}}_B=1$. The classical  action
$I_E[\phi]$  is defined on a Euclidean space with Euclidean time $\tau$ compactified
on a circle of
length $T^{-1}$.
To be more specific suppose that the partition of the space
is done orthogonally to one of the spatial coordinates, say $x^1$. Then regions $A$ and $B$ can be
defined, respectively, as $-L/2<x^1<0$ and $0<x^1<L/2$, the separating surface is
located at $x^1=0$. Relative sizes of $A$ and $B$
can be arbitrary in general.
The integration in
(\ref{12}) goes over field configurations $\phi$ with the boundary conditions
$$
\phi(x^1,...,x^{D-1},\tau=0)=\phi_B(x^1,...,x^{D-1})~~,
$$
$$
\phi(x^1,...,x^{D-1},\tau=T^{-1})=\phi'_B(x^1,...,x^{D-1})~~,
$$
where $x^1$ takes values in the interval $(0,L/2)$.

\begin{figure}[h]
\begin{center}
\includegraphics[height=5.5cm,width=5cm]{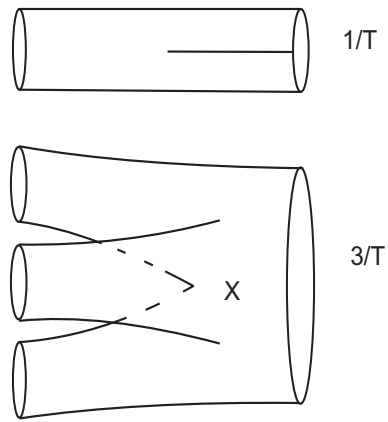}
\caption{\small{The upper picture shows ${\cal M}'_1$ in two dimensions.
It is a cylinder with the circumference length $T^{-1}$
and a cut along the axis. The space ${\cal M}_3$ is schematically drawn on the lower
picture. It is obtained
by gluing along the cuts of 3 copies of ${\cal M}'_1$. The circumference length of the right boundary of
${\cal M}_3$ is $3T^{-1}$.
The cuts meet at the point $X$ which is a conical singularity.}}
\label{f1}
\end{center}
\end{figure}

Let us denote by ${\cal M}'_1$ the space where the field configurations $\phi$ are given.
It is easy to see how ${\cal M}'_1$ looks
in two dimensions. It is a cylinder (see the upper picture
on Fig. \ref{f1}) with coordinates
$(x^1,\tau)$ and a
cut parallel to its axis, the length of the cut is equal to the length of the interval $B$.
The field takes values $\phi_B$, $\phi_B'$ on the lower and upper sides of the cut, respectively.
In higher dimensions the cut is a $(D-2)$-dimensional hyperplane.

If the parameter $n$ in (\ref{11}) is positive and integer the matrix elements of
the operator $\hat{{\rho}}_B^{~n}$ can be represented by integral
(\ref{12}) where field variables are defined on a space ${\cal M}'_n$ which is obtained
by gluing $n$ copies of ${\cal M}'_1$ along the cuts.
${\cal M}'_n$ still has a single cut which disappears  when one takes the
trace, $\mbox{Tr}_B~\hat{{\rho}}_B^{~n}$. We denote the space obtained as
a result of this procedure as ${\cal M}_n$. The lower picture on
Fig. \ref{f1} shows ${\cal M}_3$ in two dimensions.

${\cal M}_n$ is locally flat but has a non-trivial topology.
Its boundary in two dimensions
consists of $n$ circles of the length $T^{-1}$ each and a circle of the length
$nT^{-1}$.
The
important property is that ${\cal M}_n$ has a singularity at
a hyper-surface where all $n$ cuts meet. For $n\geq 2$ each point on this hyper-surface
is the conical singularity because a unit circle around it has the circumference
length equal $2\pi n$.

The singular behaviour of ${\cal M}_n$ is easy to see
in two dimensions. ${\cal M}_n$ has the same topology as a disk with $n$ holes.
Thus, its Euler characteristic,
$$
\chi[{\cal M}_n]={1 \over 4\pi}\left[\int_{{\cal M}_n}R+
2\int_{\partial {\cal M}_n} k\right]~~,
$$
equals $1-n$. The boundary $\partial {\cal M}_n$ of ${\cal M}_n$
consists of $n+1$ circles, each of which has a vanishing extrinsic curvature $k=0$.
Therefore, the integral of the scalar curvature $R$ for ${\cal M}_n$
should be non-trivial for $n> 1$,
$$
\int_{{\cal M}_n}R=4\pi(1-n)~~.
$$
The non-zero value of the integral is ensured by the conical singularity.
If this point is located at $x=0$ the curvature is the distribution
$R=4\pi(1-n)\delta^{(2)}(x)$, see, e.g. \cite{FS1}.

Let us introduce the parameter $\beta=2\pi n$ and
rewrite (\ref{11}) in another form following from
(\ref{12}),
\begin{equation}\label{13}
S_B=\lim_{\beta \rightarrow 2\pi}~ \left(\beta {\partial \over
\partial \beta}-1\right) \Gamma(\beta)~~~.
\end{equation}
$\Gamma(\beta)$ is the effective action defined by the path integral
\begin{equation}\label{14}
e^{-\Gamma(\beta)}=\int [D\tilde{\phi}]~e^{-I_E[\phi,\beta]}~~~,
\end{equation}
where field variables are given on ${\cal M}_n$. $I_E[\phi,\beta]$ is the
classical action on ${\cal M}_n$.
The normalization of the integral (\ref{14}) is fixed by the condition
$\Gamma(2\pi)=-\ln{\cal N}$.

The operation with the parameter $\beta$ in (\ref{13}) should be understood in the following way:
one first computes $\Gamma(\beta)$ for $\beta=2\pi n$, and then replaces
$\beta$ with a continuous parameter. This can be done
even if ${\cal M}_n$ itself cannot be defined at arbitrary $\beta$.

If one defines a "free energy"
$F(\beta)=\Gamma(\beta)/\beta$ and interpretes $\beta$ as an inverse temperature,
equation (\ref{13}) formally coincides with the definition of the entropy in statistical
mechanics. Note, that $\beta^{-1}$ is a "geometrical temperature" which
should not be confused with the physical temperature $T$.

\subsection{Spectral geometry on ${\cal M}_n$}

The expression of the entanglement entropy in terms of the
effective action has a number of advantages. It enables one to
reformulate the problem on the geometrical language and use
powerful methods of the spectral geometry to study the form of the entropy.

Consider a Laplace operator $\Delta=-\nabla^2$ on ${\cal M}_n$ and the trace of
its heat kernel, $\mbox{Tr}~e^{-t\Delta}$, where $t$ is a positive parameter.
The spectral geometry relates the asymptotic of the trace at small $t$ to
integral geometrical characteristics of ${\cal M}_n$,
\begin{equation}\label{as}
\mbox{Tr}~e^{-t\Delta}\sim {1 \over (4\pi t)^{D/2}}\sum_{n} a_n t^n~~~,
\end{equation}
where $n$ takes non-negative integer and half-integer values. The coefficients $a_n$ are
expressed in terms of the powers of the Riemann tensor and its derivatives,
and include boundary terms (see, e.g., \cite{Vass}). The coefficients
also depend on the singularities of the background manifold such as jumps of the curvature,
sharp corners, and so on. The effect of conical singularities appears starting
with the coefficient
$a_1$. For scalar and spinor Laplacians in two dimensions \cite{Furs:94b}
\begin{equation}\label{16}
a_1= {\pi \over 3}\left({2\pi \over \beta}- {\beta \over
2\pi}\right) +\frac 13\int_{\partial {\cal M}} k~~~.
\end{equation}
For spaces ${\cal M}_n$ like those shown on Fig. \ref{f1}
the extrinsic curvature for each
element of the boundary
is zero and $a_1$ is determined only by
the conical singularity. Other non-vanishing coefficients in the asymptotic expansion
for the heat kernel on ${\cal M}_n$ in $D=2$ are
\begin{equation}\label{16.2}
a_0=l \mbox{vol}({\cal M}_n)=l{nL \over T}=l{\beta L \over 2\pi T}~~~,
\end{equation}
\begin{equation}\label{16.3}
a_{1/2}=\sqrt{4\pi} b_{1/2}^{\pm} \int_{\partial {\cal M}_n}=\sqrt{4\pi} b_{1/2}^{\pm}
{2n  \over T}=
\sqrt{4\pi} b_{1/2}^{\pm}{\beta  \over \pi T}~~~.
\end{equation}
For scalars and spinors $l=1$  and $l=2$, respectively.
The numerical coefficients $b_{1/2}^{+}$ and $b_{1/2}^{-}=-b_{1/2}^{+}$ correspond,
respectively, to the Dirichlet  and Neumann conditions on the boundaries of ${\cal M}_n$.
For the scalar Laplacian
$b^{+}_1=-1/4$ and $b^{+}_1=-1/2$  for the spinor Laplacian.
The non-trivial dependence on $\beta$ in the first term in the
r.h.s. of (\ref{16}) is the direct consequence of the conical
singularity.

\begin{figure}[h]
\begin{center}
\includegraphics[height=4.5cm,width=4.5cm]{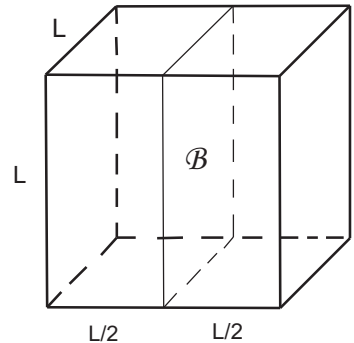}
\caption{\small{This figure shows division of a cube by a plane $\cal B$.
The trace is taken over the states of a QFT in one of the halfs of the cube.}}
\label{f4}
\end{center}
\end{figure}

Consider now QFT in a $(D-1)$-dimensional cube with the Dirichlet or Neumann boundary conditions
on its faces. Let $L$ be the length of the edge of the cube.
Suppose we are interested in the entanglement entropy
associated with the partition of the cube by a plane ${\cal B}$ which is orthogonal
to one of the faces and divides the cube into two equal parts, as is shown on Fig. \ref{f4}.
Let us denote by ${\cal M}^{(D)}_n$ a $D$--dimensional manifold which appears
in calculation of the entanglement entropy in terms of the effective action
along the lines explained in Section II.B.

In two dimensions ${\cal M}^{(2)}_n$
is shown on Fig. \ref{f1}. If $D>2$ the  manifold
${\cal M}^{(D)}_n$ is the space product of ${\cal M}^{(2)}_n$
and $D-2$ intervals of the length $L$.
For a scalar Laplace operator $\Delta^{(D)}$
on ${\cal M}^{(D)}_n$ one can write
\begin{equation}\label{heat-D}
\mbox{Tr}~e^{-t\Delta^{(D)}}=(K_L(t))^{D-2}~\mbox{Tr}~e^{-t\Delta^{(2)}}~~~,
\end{equation}
where $K_L(t)$ is the trace of the heat kernel of the Laplace operator on the interval \cite{Vass}.
The following asymptotics  hold at small $t$ up to exponentially small
terms
\begin{equation}\label{as-2}
\mbox{Tr}~e^{-t\Delta^{(2)}} \sim {1 \over  4\pi t}(a_0+a_{1/2}\sqrt{t}+a_1 t)~~~,
\end{equation}
\begin{equation}\label{as-L}
K_L(t)\sim {1 \over \sqrt{4\pi t}}(L+2b_{1/2}^{\pm} \sqrt{4\pi t})~~~,
\end{equation}
where $a_n$ are given in (\ref{16})--(\ref{16.3}). Therefore,
\begin{equation}\label{heat-D-as}
\mbox{Tr}~e^{-t\Delta^{(D)}}\sim {1 \over (4\pi t)^{D/2}}\sum_{n=0}^{D/2} a_n^{(D)} t^n~~~,
\end{equation}
where coefficients $a_n^{(D)}$ can be found from (\ref{as-2}), (\ref{as-L}).
One can check with the help of (\ref{16})--(\ref{16.3})
that the first coefficients take the form
\begin{equation}\label{a0}
a_0^{(D)}={\beta \over 2\pi T}L^{D-1}=\mbox{vol}({\cal M}^{(D)}_n)~~~,
\end{equation}
\begin{equation}\label{a1/2}
a_{1/2}^{(D)}
=\sqrt{4\pi} b_{1/2}^{\pm}{\beta  \over \pi T} (D-1)L^{D-2}
=\sqrt{4\pi} b_{1/2}^{\pm} \int_{\partial {\cal M}^{(D)}_n}~~~,
\end{equation}
$$
a_1^{(D)}={\pi \over 3}\left({2\pi \over \beta}- {\beta \over
2\pi}\right)L^{D-2} +
$$
\begin{equation}\label{a1}
(\sqrt{4\pi} b_{1/2}^{\pm})^2{\beta  \over \pi T}
(D-1)(D-2)L^{D-3} ~~~,
\end{equation}
$$
a_{3/2}^{(D)}={\pi \over 3}\left({2\pi \over \beta}- {\beta \over
2\pi}\right)(\sqrt{4\pi} b_{1/2}^{\pm})2(D-2)L^{D-3}+
$$
\begin{equation}\label{a3/2}
(\sqrt{4\pi} b_{1/2}^{\pm})^3{\beta  \over \pi T}
2^2 {(D-1)! \over 3! (D-4)!}L^{D-4} ~~~,
\end{equation}
$$
a_{2}^{(D)}={\pi \over 3}\left({2\pi \over \beta}- {\beta \over
2\pi}\right)(\sqrt{4\pi} b_{1/2}^{\pm})^2 2(D-2)(D-3)L^{D-4}+
$$
\begin{equation}\label{a2}
(\sqrt{4\pi} b_{1/2}^{\pm})^4{\beta  \over \pi T}
2^3 {(D-1)! \over 4! (D-5)!} L^{D-5} ~~~.
\end{equation}
The heat kernel coefficients for the spinor Laplacian have a similar structure and can be
found in analogous way.

Although our computations have been done in a simple setting, the results can be easily
generalized when ${\cal M}^{(D)}_n$ is  an arbitrary curved manifold.
We are particularly interested in the geometrical form of contributions
to the heat kernel coefficients from the conical singularities because they yield
a non-trivial value of the entanglement entropy.
These contributions depend on  the factor $((2\pi / \beta)- (\beta/
2\pi))$ and appear in the coefficients $a_n$ with $n\geq 1$. The geometrical
structures related to conical singularities in  the coefficients $a_1$, $a_{3/2}$, and  $a_2$
are, respectively,
\begin{equation}\label{a1-2}
L^{D-2}=\int_{\cal B}~\equiv {\cal A}~~~,
\end{equation}
\begin{equation}\label{a3/2-2}
2(D-2)L^{D-3}=\int_{\partial {\cal B}}\equiv {\cal L}~~~,
\end{equation}
\begin{equation}\label{a2-2}
2(D-2)(D-3)L^{D-4}=\int_{\mbox{\tiny{sing}}~\partial {\cal B}} \equiv {\cal C}~~~.
\end{equation}
The separating surface $\cal B$ is a $(D-2)$--dimensional manifold.
In general, it may have a boundary $\partial {\cal B}$ and the boundary may consist of
different hypersurfaces where the curvature of $\partial {\cal B}$ has jumps. Integrals in the right hand sides
of (\ref{a3/2-2}) and (\ref{a2-2}) are taken, respectively, over the boundary $\partial {\cal B}$
and codimension 1 singular hypersurfaces inside $\partial {\cal B}$.
In the considered example
of the theory in the hypercube,
$\cal B$ is a square if $D=4$ (see Fig. \ref{f4}).
It has the area ${\cal A}=L^2$ if the edge length is $L$. Its boundary
$\partial {\cal B}$ has the length $4L$ and consists of $4$ singular points, the vertexes
of the square. If $\cal B$ is a $D-2$-dimensional hypercube,
$2(D-2)$ is the number of its "faces"  and $2(D-2)(D-3)$ is the number of its "edges".

\begin{figure}[h]
\begin{center}
\includegraphics[height=3.5cm,width=7cm]{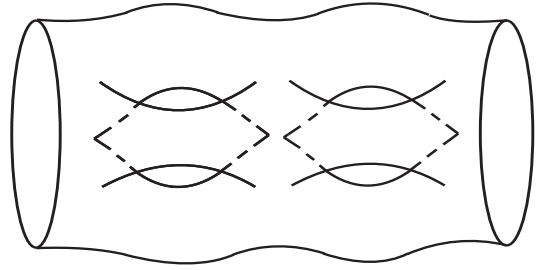}
\caption{\small{The manifold ${\cal M}^{(2)}_3$ is shown for computing the
entanglement entropy $S_B$ at a finite temperature when the region $A$ consists
of  two internal disjoint intervals.
${\cal M}^{(2)}_3$  is obtained by gluing 4 copies of the manifold
shown on Fig. \ref{f1}}. ${\cal M}^{(2)}_3$ has  4 conical singularities
at branch points.}
\label{f3}
\end{center}
\end{figure}

\subsection{Effective action and the entropy}

Let us show now how the geometrical structure of the entanglement entropy
can be established by using the effective action.
We consider, as an example, a free scalar field theory in a cube divided into two parts,
Fig. \ref{f4}. The effective action for a massive scalar field can be defined as follows
\begin{equation}\label{action}
\Gamma=\frac 12 \ln \det((\Delta+m^2)\mu^{-2})
\equiv -\frac 12 \int_{\varrho^2}^\infty {dt \over t}\mbox{Tr}~e^{-t\Delta} e^{-t m^2}~~,
\end{equation}
where $\mu$ is a dimensional parameter. The right hand side of (\ref{action})
gives the effective action in the Schwinger-De Witt representation. The parameter
$\varrho$ is a ultraviolet cutoff which has a dimensionality of a length.

It is assumed that the correlation length $m^{-1}$ of the theory
is much larger than $\varrho$. Let us also assume that $m^{-1}$
is much smaller than the size of the system $L$. In this case one can use
for $\Gamma$ an approximation  which can  be obtained when the heat kernel is replaced by its
asymptotic  (\ref{heat-D-as}).

Here and in what follows we use instead $S_B$ the notation $S$ because in the considered
examples the regions $A$ and $B$ are identical.

\bigskip
{\it 1. D=2}
\bigskip

\noindent
In two-dimensional theories
$$
\Gamma\sim -{1 \over 8\pi}\left[a_0^{(2)}\varrho^{-2}+2a_{1/2}^{(2)}\varrho^{-1}\right.
$$
\begin{equation}\label{action-2}
\left.
+(m^2 a_0^{(2)}-a_1^{(2)})\ln (m^2\varrho^2)-(a_0^{(2)}m^2-a_{1/2}^{(2)}m\sqrt{\pi}) \right]~~.
\end{equation}
In this expression we drop all terms which vanish in the limit $\varrho\rightarrow 0$.
To find the entropy one can apply (\ref{13}) to (\ref{action-2}) and use (\ref{a0})--(\ref{a1}).
It yields
\begin{equation}\label{s2}
S\sim -{1 \over 12}\ln (m^2\varrho^2)~~~.
\end{equation}
If $\varrho$ is identified with the lattice spacing,
this result coincides with computations in spin chains in the near critical regime,
see (\ref{r4}). The central charge in the given case is $c=1$.
Formula (\ref{s2}) is obtained  for the partition of the
interval into two parts. It can be easily generalized for more complicated cases
when the regions $A$ and $B$ consist of several disjoint intervals.
The effective action in this case is defined on a 2D manifold ${\cal M}^{(2)}_n$
with some number of handles. Figure \ref{f3} shows ${\cal M}^{(2)}_3$ for
tracing the degrees of freedom in two disjoint internal intervals.
As is easy to understand the number of conical singularities ${\cal N}_c$ of ${\cal M}^{(2)}_n$
equals the number of internal points of the intervals.
Each conical singularity yields an independent contribution to the entropy. Thus, in
general case the result in the right hand side of (\ref{s2}) has to be multiplied
by ${\cal N}_c$ \cite{CaCa:04}.

\bigskip
{\it 2. D=3}
\bigskip

\noindent
In three-dimensional theories
$$
\Gamma\sim -{1 \over 16\pi^{3/2}}\left[\frac 23 a_0^{(3)}\varrho^{-3}+
a_{1/2}^{(3)}\varrho^{-2}+\right.
$$
$$
2(a_1^{(3)}-m^2 a_0^{(3)})\varrho^{-1}
-(a_{3/2}^{(3)}-m^2 a_{1/2}^{(3)})\ln (m^2\varrho^2)+
$$
\begin{equation}\label{action-3}
\left. \left({2 \over 3\sqrt\pi}a_0^{(3)} m^3-a_{1/2}^{(3)}m^2-\sqrt{\pi}a_1^{(3)} m\right)
\right]~~
\end{equation}
and the entanglement entropy takes the form
\begin{equation}\label{s3}
S\sim {1 \over 12\sqrt{\pi}}\left[\left({1 \over \varrho}-{\sqrt{\pi} \over 2}m\right){\cal A} -
b_{1/2}^{\pm}\ln (m^2\varrho^2)~{\cal L}\right]~~~,
\end{equation}
where we used (\ref{a1-2}),(\ref{a3/2-2}).
For the theory in the square ${\cal A}=L$, ${\cal L}=2$.
The term in (\ref{s3}) which is proportional to ${\cal L}$
 has the following property: it does not change
under continuous deformation of the separation surface $\cal B$.
This property is in agreement with a more
general behaviour of the entanglement
entropy in three-dimensional theories which has a topological part arising from short wavelength
modes localized near the boundary of $\cal B$, see \cite{KiPr}.

\bigskip
{\it 3. D=4}
\bigskip

\noindent
In four-dimensional theories
$$
\Gamma\sim -{1 \over 32\pi^2}\left[\frac 12 a_0^{(4)}\varrho^{-4}+
\frac 23 a_{1/2}^{(4)}\varrho^{-3}+\right.
$$
$$
(a_1^{(4)}-m^2 a_0^{(4)})\varrho^{-2}+2(a_{3/2}^{(4)}-m^2 a_{1/2}^{(4)})\varrho^{-1}-
$$
$$
(a_2^{(4)}-m^2a_1^{(4)}+{m^4 \over 2}a_0^{(4)})\ln (m^2\varrho^2)+
$$
\begin{equation}\label{action-4}
\left.
\left(\frac 34 a_0^{(4)}m^4 + {2 \over 3\sqrt{\pi}}a_{1/2}^{(4)}m^3-
a_1^{(4)}m^2+\sqrt{\pi}a_{3/2}^{(4)} m\right)
 \right] ~~.
\end{equation}
Equations (\ref{a0})--(\ref{a2-2}) give the following expression for the
entanglement entropy
$$
S\sim {1 \over 48\pi}\left[\left({1 \over \varrho^2}+m^2(\ln(m^2\varrho^2)-1)
\right){\cal A} -\right.
$$
\begin{equation}\label{s4}
\left.
4\sqrt{\pi} b_{1/2}^{\pm}~\left({2 \over \varrho}-\sqrt{\pi} m\right) ~{\cal L}
-{\pi \over 4}~\ln (m^2\varrho^2)~{\cal C} \right]~~~.
\end{equation}
For a theory in the cube, ${\cal L}=4L$, ${\cal C}=4$. In general,
${\cal C}$ is the number of singular points of the boundary $\partial {\cal B}$
of the separation
surface. The term in (\ref{s4}) which is proportional to ${\cal C}$ does not
change under the continuous deformation of $\partial {\cal B}$. Therefore, it is
a topological term similar to the subleading term in the entropy in three dimensions.

\bigskip

Expressions (\ref{s3}), (\ref{s4}) hold for relativistic theories. In general,
dispersion relations in condensed matter systems
are not Lorentz invariant and are not linear. As a result, the leading
term in the ground state entanglement entropy may have a more complicated
dependence on the area of the partition surface $\cal A$ than in (\ref{s3}), (\ref{s4}).
In particular, an additional logarithmic coefficient pointed out in \cite{GiKl}--\cite{Barthel}
may appear in the scaling of the entropy.

\subsection{Ground state entanglement at critical points}

So far we assumed that the system has a non-vanishing correlation length which is small
compared to the size of the system. At critical points the correlation length becomes infinite and
approximations considered above do not hold. In general, the effective action in this
regime is non-local and its computation is non-trivial.

The effective action $\Gamma$ is the sum
$$
\Gamma=\Gamma_{\mbox{\tiny{div}}}+\Gamma_{\mbox{\tiny{ren}}}~~.
$$
$\Gamma_{\mbox{\tiny{div}}}$ diverges
in the limit of vanishing cutoff $\varrho$ while the "renormalized" part
$\Gamma_{\mbox{\tiny{ren}}}$ is finite at $\varrho=0$. The structure of divergences $\Gamma_{\mbox{\tiny{div}}}$
can  be found from (\ref{action-2}), (\ref{action-3}),(\ref{action-4}). To obtain $\Gamma_{\mbox{\tiny{div}}}$
in the massless theory one has to leave in (\ref{action-2}), (\ref{action-3}),(\ref{action-4})
only terms depending on $\varrho$ and take the limit $m=0$.

For critical systems whose size is
parametrized by a single parameter the renormalized part can be found by using
anomalous scaling properties of $\Gamma_{\mbox{\tiny{ren}}}$.
Let us consider as before a free scalar field theory in a cube of the edge size $L$.
The effective action is $\Gamma=\frac 12 \ln \det \Delta \mu^{-2}$.
If one changes the operator $\Delta$ to $\Delta'=e^{2\sigma} \Delta$, where
$\sigma$ is a constant, the renormalized part
of the action for  $\Delta'$ can be written as
\begin{equation}\label{anom-1}
\Gamma'_{\mbox{\tiny{ren}}}=\Gamma_{\mbox{\tiny{ren}}}+
\sigma {a_{n=D/2}^{(D)} \over (4\pi)^{D/2}}~~~.
\end{equation}
This result can be easily obtained in the dimensional regularization \cite{DoSc:90}.
The last term in the right hand side of this equation is known as conformal anomaly.
In the considered case the action is a function $\Gamma_{\mbox{\tiny{ren}}}(\beta,L,T,\mu)$.
By using (\ref{anom-1}) one can write
$$
\Gamma_{\mbox{\tiny{ren}}}(\beta,L,T,\mu)=
$$
\begin{equation}\label{anom-2}
\Gamma_{\mbox{\tiny{ren}}}(\beta,L_0,(LT)/L_0,\mu)-
{a_{n=D/2}^{(D)} \over (4\pi)^{D/2}}\ln {L \over L_0}~~~.
\end{equation}
Analogously, the entanglement entropy can be written as
\begin{equation}\label{stot}
S=S_{\mbox{\tiny{div}}}+S_{\mbox{\tiny{ren}}}~~~,
\end{equation}
where $S_{\mbox{\tiny{div}}}$ and $S_{\mbox{\tiny{ren}}}$ are divergent and finite parts
which are calculated, respectively, from $\Gamma_{\mbox{\tiny{div}}}$ and
$\Gamma_{\mbox{\tiny{ren}}}$ by using equation (\ref{13}).
According to (\ref{anom-2}), the renormalized part can be written as
\begin{equation}\label{sren}
S_{\mbox{\tiny{ren}}}(L,T,\mu)=S_{\mbox{\tiny{ren}}}
(L_0,(LT)/L_0,\mu)+{s_{n=D/2}^{(D)} \over (4\pi)^{D/2}}\ln {L \over L_0}~~~,
\end{equation}
\begin{equation}\label{sD/2}
s_{n=D/2}^{(D)}=-\lim_{\beta \rightarrow 2\pi}~ \left(\beta {\partial \over
\partial \beta}-1\right)a_{n=D/2}^{(D)}~~~.
\end{equation}
The part of the entropy $S_{\mbox{\tiny{ren}}}(L_0,(LT)/L_0,\mu)$ is determined by
$\Gamma_{\mbox{\tiny{ren}}}(\beta,L_0,(LT)/L_0,\mu)$.

To get the entropy in the ground state one has to go to the limit $T\rightarrow 0$.
It is known from the lattice computations that the entanglement entropy is
finite at zero temperature. This property implies that
$S_{\mbox{\tiny{ren}}}(L_0,(LT)/L_0,\mu)$ in the zero temperature limit should be a finite
constant, $C(L_0,\mu)$, which does not depend on the size $L$.
Therefore,
\begin{equation}\label{sren0}
S_{\mbox{\tiny{ren}}}(L,T=0,\mu)=
{s_{n=D/2}^{(D)} \over (4\pi)^{D/2}}\ln {L \over L_0}+C(L_0,\mu)~~~.
\end{equation}
By using equations (\ref{a1})--(\ref{a2}), (\ref{s2}), (\ref{s3}), (\ref{s4}), (\ref{sD/2}) and
(\ref{sren0}) one finds the ground state entanglement entropy of a massless scalar
field for a theory in a cube in dimensions $D=2,~3,~4$, respectively,
\begin{equation}\label{s2-0}
S\sim {1 \over 6}\ln (L/\varrho)~~~,
\end{equation}
\begin{equation}\label{s3-0}
S\sim {1 \over 12\sqrt{\pi}}\left[\varrho^{-1}{\cal A}
+2b_{1/2}^{\pm}\ln (L/\varrho)~{\cal L}\right]~~~,
\end{equation}
\begin{equation}\label{s4-0}
S\sim {1 \over 48\pi}\left[\varrho^{-2}~ {\cal A}+
4\sqrt{\pi} b_{1/2}^{\pm}~\varrho^{-1} ~{\cal L}
+{\pi \over 2}~\ln (L/ \varrho)~{\cal C} \right]~~~,
\end{equation}
where we omitted constants which do not depend on the parameters of the system.
By comparing (\ref{s2-0})--(\ref{s4-0}) and (\ref{s2}), (\ref{s3}), (\ref{s4}) one can
conclude that the size of the system $L$ in the massless theory plays the role of the
infrared cutoff analogous to the inverse mass.
The result for the entropy in two dimensions is in agreement with the entanglement entropy
in the critical spin chains, see (\ref{r3}) and the earlier computations of the entropy
in \cite{HLW}.

Although formulae (\ref{s2-0})--(\ref{s4-0}) are obtained for the theory of a free field,
they should preserve their form in the presence of interactions. The logarithmic terms
related to the conformal anomaly should not receive corrections.

\subsection{High temperature limit}
\bigskip

The asymptotic of the entanglement entropy can be also found
in the opposite limit, $TL \gg 1$, $T\gg m$. In this limit the main
contribution comes from the modes with the wave length of the order of $T^{-1}$.
The theory in this case is extensive. Consider first the theory
on an interval $L$ divided into two parts.
The corresponding background manifold ${\cal M}^{(2)}_n$ is shown on Fig. \ref{f1}.
As before, let $\Gamma_{\mbox{\tiny{ren}}}(\beta,L,T,\mu)$ be a renormalized
action on ${\cal M}^{(2)}_n$ ($\beta=2\pi n$). Let
$\tilde{\Gamma}(\beta_0,L/2,T,\mu)$, $\beta_0=2\pi$,  be the action
on the cylinder of the length $L/2$. The space
${\cal M}^{(2)}_n$ can be obtained by gluing $n$ cylinders of length $L/2$ and
the circumference length $T^{-1}$ with the cylinder of length $L/2$ and
the circumference length $nT^{-1}$. Because the theory in the high temperature limit
is extensive the action  can be written as
$$
\Gamma_{\mbox{\tiny{ren}}}(\beta,L,T,\mu)\sim
$$
\begin{equation}\label{ht-1}
n\tilde{\Gamma}(\beta_0,L/2,T,\mu)
+\tilde{\Gamma}(\beta_0,L/2,T/n,\mu)~~~.
\end{equation}
This relation also holds in higher dimensions.
It is not difficult to show by knowing the free energy  of a one-dimensional
gas on an interval that at high temperatures
$$
\tilde{\Gamma}(\beta_0,L/2,T,\mu)\sim -{\pi \over 6} (L/2)T~~~.
$$
(The effective action is a free energy divided by temperature.)
Therefore,
\begin{equation}\label{ht-2}
\Gamma_{\mbox{\tiny{ren}}}(\beta,L,T,\mu)\sim
-\left({\beta \over 2\pi}+{2\pi \over \beta}\right){\pi \over 6} (L/2)T
~~~.
\end{equation}
At $\beta=\beta_0$ this result coincides with the effective action on the cylinder of the length $L$.
The divergent part of the effective action in this limit can be neglected and the
entropy can be determined from (\ref{ht-2}) with the help of (\ref{13}),
\begin{equation}\label{ht-3}
S(L,T,\mu)\sim
{\pi \over 3} (L/2)T
~~~.
\end{equation}
The entanglement entropy in the high temperature limit
coincides with the usual statistical entropy of the system of the length $L/2$.
A similar result was found in \cite{CaCa:04}.

In three and four-dimensional spacetimes the effective action for a
scalar theory in a volume $V$ at high temperatures behaves, respectively, as
$-\zeta(3) VT^2/(2\pi)$,
$-\pi^2 VT^3/(90)$, see \cite{DoSc:88}. From (\ref{ht-1}) one then finds
high temperature entanglement entropy for partition of a square or a cube
into two parts
\begin{equation}\label{ht-4}
S(L,T,\mu)\sim {3\zeta(3) \over 2\pi} VT^2
~~~,
\end{equation}
\begin{equation}\label{ht-5}
S(L,T,\mu)\sim
{2\pi^2 \over 45} VT^3
~~~.
\end{equation}
Here $\zeta(x)$ is the Riemann zeta function and $V=L^{D-1}/2$. In general case,
$V$ in (\ref{ht-4}), (\ref{ht-5}) is the volume of that
region where the entanglement entropy is calculated.
Note that in this regime the entanglement entropy coincides with the usual statistical entropy
in the volume $V$, in agreement with extensive properties of the system at high temperatures.

\section{Fundamental entanglement and induced gravity}
\subsection{Induced Newton coupling}

According to Sakharov's  idea \cite{Sakh}, the classical
Einstein-Hilbert gravitational action
can be entirely induced by quantum effects of matter fields on a curved
background. As an example, consider a model of a free scalar field of the mass $m$
with an ultraviolet cutoff $\varrho$. The model is defined
on a curved manifold $\cal M$ with metric  $g_{\mu\nu}$.  To keep connection with
finite-temperature theories we assume that $g_{\mu\nu}$ has the Euclidean signature.
For simplicity we also assume that $\cal M$ is a compact closed manifold.
The effective action of the field can be found by using
(\ref{action}). The heat kernel operator now has the following asymptotic (see e.g.
\cite{Vass})
\begin{equation}\label{in-1}
\mbox{Tr}~e^{-t\Delta}\sim {1 \over (4\pi t)^{D/2}}\sum_{n} A_n^{(D)} t^n~~~,
\end{equation}
\begin{equation}\label{in-2}
A_0^{(D)}=\int_{\cal M}\sqrt{g} d^Dx~~,~~A_1^{(D)}\simeq \frac 16 \int_{\cal M}\sqrt{g} d^Dx ~R~~~,
\end{equation}
where $g=\det g_{\mu\nu}$, $R$ is the scalar curvature of
$\cal M$, $n$ takes non-negative integer values.
The higher coefficients are integrals of powers of the Riemann tensor and its derivatives.

We consider theory in $D=4$. If the curvature radius
of $\cal M$ is much larger than $m^{-1}$ on can proceed as in Section II.D
and get
$$
\Gamma\sim -{1 \over 32\pi^2}\left[\frac 12 A_0^{(4)}\varrho^{-4}+
(A_1^{(4)}-m^2 A_0^{(4)})\varrho^{-2}-\right.
$$
$$
(A_2^{(4)}-m^2A_1^{(4)}+{m^4 \over 2}A_0^{(4)})\ln (m^2\varrho^2)-
$$
$$
\left.
\left(\frac 34 A_0^{(4)}-
A_1^{(4)}m^2\right)\right]\simeq
$$
\begin{equation}\label{in-3}
-{1 \over 16\pi G} \int_{\cal M}\sqrt{g} d^Dx ~(R-2\Lambda)~~.
\end{equation}
In the last line all terms related to $A_2^{(4)}$ were omitted.
The right hand side of (\ref{in-3}) coincides with the classical Einstein-Hilbert
action where  $G$ and $\Lambda$ are the induced Newton and cosmological
constants, correspondingly,
\begin{equation}\label{27}
{1 \over G}={1  \over 12\pi} \left({1 \over  \varrho^2} +
m^2 (\ln (m^2 \varrho^2)-1)\right)~~~,
\end{equation}
\begin{equation}\label{in-4}
{\Lambda \over G}=-{1  \over 8\pi} \left({1 \over  \varrho^4} -
{2m^2 \over  \varrho^2}+
m^4 \left(\ln (m^2 \varrho^2)-\frac 32\right)\right)~~~.
\end{equation}
The constant $G$ has the correct order of magnitude if the cutoff parameter $\varrho$
is of the order of the Planck length.

By comparing (\ref{27}) with (\ref{s4}) we conclude that, in accord with the hypothesis (\ref{i3}),
the induced Newton constant in this simple model and the entanglement
entropy of the system in the flat space-time in a cube
are related as
\begin{equation}\label{in-5}
{1 \over G}=4\lim_{L \rightarrow \infty}{S \over {\cal A}}~~~.
\end{equation}
Here ${\cal A}=L^2$ is the area of the separation surface.
The result holds when the size of the cube
is sufficiently large.

The reason why (\ref{in-5}) takes place is
because the conical space possesses a curvature concentrated at the tip.
Let us consider the heat kernel coefficient $a_1^{(D)}$, Eq. (\ref{a1}), for the theory on
${\cal M}^{(D)}_n$ and compare it with $A_1^{(D)}$ in (\ref{in-2}).
If $\beta$ in (\ref{a1})  is close to
$2\pi$ then
\begin{equation}\label{in-6}
a_1^{(D)}\simeq {1 \over 6}2(2\pi-\beta){\cal A}+\mbox{b.t.}=
\frac 16 \int_{{\cal M}^{(D)}_n} ~R_c+\mbox{b.t.}~~~,
\end{equation}
where $R_c=2(2\pi-\beta)\delta^2(x)$ is the
"curvature" of ${\cal M}^{(D)}_n$ concentrated at  conical singularities
on the hypersurface $x=0$. Notation "b.t." stands for boundary terms which are irrelevant
for the computation of the entropy. Therefore, the coefficients
$a_1^{(D)}$ and $A_1^{(D)}$ have the same geometrical form in this limit.

The similarity of the heat kernel coefficients on smooth manifolds and manifolds with
conical singularities at small angle deficits is a quite general property. If the components
of the Riemann tensor are treated at the conical singularities as distributions
\cite{FS1} the coincidence holds also for higher
heat kernel coefficients and for different field species \cite{FS2}.

It should be noted that (\ref{in-5}) does not hold for theories with fields which
have non-minimal couplings to the curvature.
A typical example is the scalar theory with the wave operator
$\Delta-\xi R$, where $\xi$ is a constant and $R$ is the scalar curvature.
The coefficient $a_1^{(D)}$ does not depend on $\xi$ because, by the construction of
the effective action $\Gamma(\beta)$ in Section II.B,  the wave operator
on ${\cal M}^{(D)}_n$ coincides with the Laplacian $\Delta$. However, the coupling changes
the heat kernel coefficients on curved manifolds. In particular, the multiplier
$\frac 16$ in $A_1^{(D)}$ in (\ref{in-2}) is replaced by $\frac 16-\xi$.

It is still an open question whether the entanglement entropy in a physical theory
depends on the non-minimal couplings. The answer to it may be related to the
procedure of the measurements of the energy (a discussion of the topic can be found
in \cite{FF:97}).
In what follows we consider theories without non-minimal couplings.

\subsection{The gravitational coupling and analogue models of gravity}

It is natural to suggest that a fundamental gravity theory and a simple model with a cutoff
considered in the previous Section share two features:
i) in the both cases the low-energy gravity sector is a pure effective theory and equations for the metric
tensor
are determined entirely by the polarization properties of the physical vacuum;
ii) the theories do not have a problem of the ultraviolet divergences.

The string theory which is considered as one of the candidates to the
role of the fundamental theory has these features.
The second property ensures that the induced Newton coupling $G_N$ is finite and can be
expressed in terms of microscopical parameters. If the non-minimal
couplings  are absent $G_N$ can be derived from
the ground state entanglement in a flat spacetime by using (\ref{in-5}).

To get more intuition about quantum gravity it makes sense to study other models
where the effective Newton coupling can be calculated with the help of (\ref{in-5}).
One of the possibilities is to consider condensed matter systems
which are higher-dimensional generalizations of spin chains
discussed in Section II.A. The requirements to the models are:
1) the models are lattice theories;
2) they have critical points which correspond to second order phase transitions;
3) near the critical points the models are described by relativistic QFT's with
massive fields.

The cutoff in this type of models is the lattice spacing. The entanglement
entropy is finite and it can be computed numerically.
The third requirement guarantees that
in the near critical regime one can use results of Section II.D to get the leading
asymptotic of the entropy. One can use for this purpose an effective QFT
(not the underlying lattice theory itself). Then the density
of the entropy per unit area yields the induced gravitational coupling (relation (\ref{in-5})).

One of examples of such systems are higher-dimensional Ising models which
are known to be equivalent in the critical regime to scalar field theories with
self-interactions \cite{Pol:87}. The models are also interesting for other reasons.
The behavior of the most part of known  second-order phase transitions
is equivalent at the critical point to a three-dimensional (3D) Ising model \cite{Pol:87}.
There are also indications that Ising models can be represented as theories of
random (hyper) surfaces. In particular, as was conjectured in \cite{Pol:87}
near the point of the second order phase transition the 3D Ising model might be equivalent to
a non-critical fermionic string theory.

The two-dimensional Ising model is exactly solvable. The reduced density matrix for the
ground state entanglement can be diagonalized. This property significantly simplifies
computations of the entropy. In higher dimensions exactly solvable Ising models are not known.
Thus,
one has to rely on numerical methods.

\bigskip

It should be noted that the role of condensed matter systems as gravity analogs
is known for many years but in a different context: to study
quantum effects in the external gravitational fields \cite{Unruh}.
One of such effects is the Hawking radiation. Under certain
conditions sound waves in a liquid Helium or in Bose-Einstein
condensates behave as scalar excitations propagating in an
effective curved background with a metric similar to that near a
black hole horizon (see, e.g., \cite{Volovik} and references
therein).

Let us emphasize that
because we are
interested only in the behavior of the Newton coupling we do not need to introduce any metric
(or its analog) in the
condensed matter system. To determine the coupling it is sufficient to study the response
of the effective action
in flat space to introduction of conical singularity.

\section{Applications}
\subsection{RG-flow of the gravitational coupling}

The knowledge
of entanglement entropy as a function of parameters of the theory can be used to find
renormalization group (RG) evolution
of the induced Newton constant.
This information is important for understanding  the behaviour of gravitational
interactions at different scales.

Consider as an illustration a one-dimensional spin chain.
It has a  compact momentum space with the radius
$\bar{p}=2\pi/\varrho$ determined by the lattice spacing $\varrho$. The Wilson RG transformation
implies integration over high-energy
modes with momenta in the interval $t^{-1} \bar{p}\leq p \leq \bar{p}$ (with $t>1$)
followed by
rescaling, $p$ to $p'=t p$. As a result one gets a theory
with larger masses. For the Ising model (\ref{6}) with $\lambda \neq 1$
this is  equivalent to increasing the
difference $|\lambda-1|$.
The RG-transformation drives the theory away from ultraviolet fixed point $\lambda=1$.

\begin{figure}[h]
\begin{center}
\includegraphics[height=4cm,width=6cm]{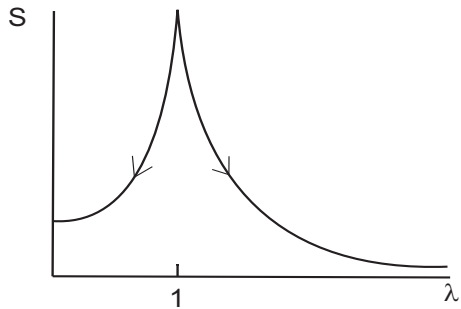}
\caption{\small{This figure is based on results of \cite{crit2}. It shows
the dependence of the entanglement entropy in the Ising
spin chain (\ref{6}) at a fixed $N$ as a function of magnetic field strength $\lambda$.
The critical point $\lambda=1$ is ultaviolet fixed point. The arrows show directions
of the RG flow from the ultraviolet to infrared regions.}}
\label{f2}
\end{center}
\end{figure}

The RG-evolution of the entanglement entropy for the Ising model
is known precisely \cite{crit2}.
The dependence of the entanglement entropy in the Ising
spin chain (\ref{6}) at a fixed $N$ as a function of magnetic field strength $\lambda$
is shown on Fig. \ref{f2}.
This behaviour  is in accord with a
general property:
in unitary theories the entanglement entropy should not increase along the RG flow
because RG transformations eliminate contribution of the high energy modes.

The fact that the entropy is not increasing does not imply the
same property for the induced coupling $G^{-1}$ defined by
(\ref{in-5}). The coupling behaves as the density of the entropy,
$G^{-1}(t)=t^{D-2} f(t)$, where $D$ is the number of
space-time dimensions and $f(t)$ is a function which has
the same RG-evolution as the entropy. In three-dimensional model discussed in Section II.D
the induced gravitational coupling defined by (\ref{in-5}) increases under RG-evolution.
On the other hand, in four-dimensional model the gravitational coupling decreases
which means that gravitational interactions get weaker in the infrared region.


\subsection{The scaling hypothesis}

The scaling hypothesis in  classical critical phenomena asserts
that the physics is determined by large scale fluctuations
which do not depend on the underlying microscopical details. For instance, in a magnetic
at temperature $T$ in an external field $h$ which
undergoes a second order phase transition the physical properties
are determined by the large domains of aligned spins.
The microscopic atomic scale does not enter  thermodynamical
relations near the critical point $T=T_c$.
Non-analytic  dependence of the physical quantities on
$|T-T_c|$ is entirely determined by a "singular part" of free energy density
$f_{\mbox{\tiny{sing}}}$.  This part has a universal scaling
$$
f_{\mbox{\tiny{sing}}}(t,h)=\xi^{-(D-1)}f_{\pm}(h\xi^{-d_h})~~~,
$$
where $\xi \sim |T-T_c|^{-\nu}$ is the correlation length  and
$d_h$ is the scaling dimension of $h$. Functions $f_{\pm}$
are universal in a sense they coincide for different systems from the same class.

Let us denote by ${\cal S}_{\mbox{\tiny{ren}}}$ the surface density of the "renormalized" part
of the entanglement entropy in the ground state, see Section II.E.
This is the part of the entropy
which remains finite in the limit $\varrho=0$. In general,
${\cal S}_{\mbox{\tiny{ren}}}$ is
a function of  external fields $h$, and other parameters, $g$,
driving the phase transition of the system at $g=g_c$.
By using analogy with the classical scaling one can conjecture
that near a critical point ${\cal S}_{\mbox{\tiny{ren}}}$ has the following
scaling \footnote{Here we follow the conjecture made in \cite{CaCa:04}. However, as distinct from
\cite{CaCa:04} we assume that (\ref{scale}) holds only for the ground state
entanglement entropy.}
\begin{equation}\label{scale}
{\cal S}_{\mbox{\tiny{ren}}}(g,h)=\xi^{-(D-2)}s_{\pm}(h\xi^{-d_h})~~~,
\end{equation}
where $\xi=|g-g_c|^{-\nu}$ is the correlation length, $d_h$
is the scaling dimension, and $s_{\pm}$ are universal functions.
It can be also suggested that the divergent
part of the entropy is analytic in $g$.

One can check the validity of these statements in scalar field models discussed
in Section II.D. In the limit when the edge size $L$ of the cube is large
one finds from (\ref{s3}), (\ref{s4}) that  $s_{\pm}=-1/(24)$ in $D=3$ and
$s_{\pm}=-1/(24\pi)\ln(m/\mu)$ in $D=4$.

In analogue models of gravity the universality conjecture (\ref{scale}) means that
the part of the induced Newton coupling (\ref{in-5}) which does not depend on the cutoff
is a universal function. This part of the coupling can be determined entirely in terms
of an effective low-energy QFT and it does not depend on the details of the underlying
microscopic theory.


\subsection{The problem of black hole entropy}

Studying entanglement in gravity analogs might be helpful for understanding the microscopical origin of
the Bekenstein-Hawking entropy $S^{BH}$ of black holes. As was discussed in Section I.B,
if the gravity is entirely induced
by some underlying degrees of freedom the  entropy of a black hole can be related
to the entanglement between observable states and
states hidden inside the horizon.
Understanding the relation between the two entropies in the framework of a local relativistic
quantum field theory is plagued by the problem of the ultraviolet divergences. The definition of
the induced Newton coupling either requires introduction of the ultraviolet cutoff or
working with a special class of ultraviolet finite theories with non-minimal
couplings \cite{FFZ:96}, \cite{FF:97}.
The presence of non-minimal couplings
makes statistical interpretation of $S^{BH}$ a difficult task.

The equality of  $S^{BH}$ to the
fundamental entanglement entropy $S$ is implied
in suggestion (\ref{i3}). One of the arguments why it should be true is that  both in
classical and in quantum theory the  Bekenstein-Hawking entropy can be derived by the conical singularity method from
the gravitational action (see e.g. \cite{FrFu:98}, \cite{FS1}),
in the same way which we employed for calculation of $S$.

If equality (\ref{i3}) holds one has a tool to learn more about
the degrees of freedom responsible for the origin of $S^{BH}$  by using gravity analogs. There are
different questions which can be addressed in different models. For instance,
can one relate these degrees of freedom in the 3D Ising model with non-critical stings,
do the near-horizon symmetries play any role in the entropy counting as was
suggested in \cite{Carlip:05}, and other questions.

\bigskip

\section{Summary}

This paper has two purposes. First, we wanted to show how the effective action
approach and the methods of spectral geometry can be applied for derivation of
the entanglement entropy in condensed matter systems in the regimes when these
systems can be described by relativistic QFT's. In particular, such regimes appear in the
vicinity of the second order phase transitions. By using the effective action approach
we established the form of the entanglement entropy  in the
presence of boundaries at zero and high temperatures.
Our new result is establishing the geometrical structure of the subleading terms in the
entropy, Eq. (\ref{main}).

Second, we showed that in the effective action approach the
black hole entropy and the entanglement entropy have the same geometrical origin, the
conical singularities of a background manifold. On this base we conjectured (and this is the
main result of the paper) that
the surface density of the fundamental entanglement entropy in the genuine quantum gravity theory
and the low-energy gravitational coupling $G_N$ are related by (\ref{i3}).
Finally, we suggested to use this relation to study the properties of $G_N$
in analogue gravity models and pointed several topics where this kind of research could
be interesting.

\noindent
\section*{Acknowledgment}\noindent
The author is grateful to V. Frolov, A. Mironov, A. Morozov, and A. Zelnikov for helpful
discussions during the preparation of this work.
This work was supported in part by the Scientific School Grant N 5332.2006.2.

\end{document}